\documentclass[11pt]{article}
\usepackage{amsmath,amssymb}
\usepackage[dvips]{graphicx}
\usepackage[abs]{overpic}
\newcommand{\ba}{\begin{alignat}{3}}

\usepackage[dvipdfm]{color}

\begin{document}


\begin{flushright}
OU-HET 658
\\
June 1, 2010
\\
arXiv:1003.1203 [hep-th]
\end{flushright}
\vskip3cm
\begin{center}
{\LARGE {\bf  Galilean Conformal Algebra 
in Two Dimensions 
\\[0.2cm]
and  Cosmological Topologically Massive Gravity}}
\vskip3cm
{\large 
{\bf Kyosuke Hotta,\,
Takahiro Kubota
 \,and\, Takahiro Nishinaka
}
}

\vskip1cm
{\it Department of Physics, Graduate School of Science, 
\\
Osaka University, Toyonaka, Osaka 560-0043, Japan}
\end{center}

\vskip1cm
\begin{abstract}
We consider a  realization of the Galilean conformal algebra (GCA) in two dimensional space-time  on the AdS boundary of a particular three dimensional gravity theory, the so-called cosmological topologically massive gravity (CTMG), which includes the gravitational Chern-Simons  term and the negative cosmological constant.
The infinite dimensional GCA in two dimensions is obtained from the Virasoro algebra for the relativistic CFT by taking a scaling  limit $t\rightarrow t$, $x\rightarrow\epsilon x$ with $\epsilon\rightarrow 0$.
The parent relativistic CFT should have left and right central charges of order $\mathcal{O}(1/\epsilon)$ but opposite in sign in the limit $\epsilon\rightarrow 0$. 
On the other hand,  by Brown-Henneaux's analysis the Virasoro algebra is realized on the boundary of AdS$_3$, but the left and right central charges are asymmetric only by the factor of the gravitational Chern-Simons  coupling $1/\mu$.
If $\mu$ behaves as of order $\mathcal{O}(\epsilon)$  
under the corresponding  limit, we have the GCA  with  non-trivial centers on AdS boundary of the bulk CTMG. 
Then we present a new entropy formula for the Galilean field theory from the bulk black hole entropy, which is a non-relativistic counterpart of the Cardy formula.
It is also discussed whether it can be reproduced by the microstate counting. 
\end{abstract}


\vfill\eject


\section{Introduction}

We are now facing a new stage for the application of the duality between bulk gravity theory and boundary quantum field theory. 
Above all, the AdS/CFT correspondence, on which much attention has been  focused in the past  decade, enables us  not only to understand  still  enigmatic quantum aspects of gravity but to give a powerful tool to investigate regions unexplored 
in the usual field theory approach.
It is certain that  the generalization of this bulk/boundary connection to  non-relativistic settings, some of which may be realized closely to our laboratory, or possibly to our real life, would provide a more useful dictionary for us to quest for the complete quantum field theory.

The Schr\"{o}dinger symmetry group  is one of the non-relativistic conformal symmetry, and  can be seen in  systems, such as cold atoms~\cite{nishida}. 
This has $SO(d,2)$ conformal symmetry in $d$  dimensions with dynamical exponent $z=2$.
The gravity dual for the non-relativistic CFT with the Schr\"{o}dinger symmetry was proposed in refs.~\cite{son, bala}, and then the non-relativistic version of the AdS/CFT correspondence has recently been explored to some extent.
In this paper, however, we focus on  another non-relativistic realization of the conformal symmetry, the so-called Galilean conformal algebra (GCA), and consider its gravity dual in the bulk.  
For the recent developments on the GCA, see the papers~\cite{gopa1}-\cite{man}.
 
Unlike the Schr\"{o}dinger case, the Galilean conformal symmetry  has dynamical exponent $z=1$ and allows the infinite dimensional algebra  in any space-time dimensions, which contains local conformal transformations, rotations and boosts.
Especially, an important thing in two dimensions is that the infinite dimensional GCA  can actually be obtained from the Virasoro algebra for the relativistic CFT$_2$ by taking a non-relativistic limit $t\rightarrow t$, $x\rightarrow\epsilon x$ with $\epsilon\rightarrow 0$~\cite{hosseiny-rouhani,gopa2}.
By the GCA construction defined below, it  turns  out that for finite GCA central charges, $C_1$ and $C_2$, the Virasoro central charges $c^\pm$ must behave as $c^{+}+c^{-}=\mathcal{O}(1)$ and 
$c^{+}-c^{-}=\mathcal{O}(1/\epsilon)$ 
under the non-relativistic limit.
In addition, for finite scaling dimensions $\Delta$ and rapidity $\xi$, which are labels specifying a primary state of the GCA, the conformal weight of the parent Virasoro algebra $h^\pm$  must also behave as 
$h^{+}+h^{-}=\mathcal{O}(1)$, and 
$h^{+}-h^{-}=\mathcal{O}(1/\epsilon)$.
The main subject of this paper is to reconstruct the GCA in two dimensions with these properties, at least, the finite central charges $(C_1,C_2)$ and the labels $(\Delta,\xi)$ from bulk gravity theory in the context of the AdS/CFT correspondence.

According to ref.~\cite{gopa1}, the GCA in $d$ dimensions is realized on the boundary of the  AdS$_{d+1}$ gravity, but this dual theory must be reduced by  a certain  limit just like the usual non-relativistic gravity theory.
In the asymptotically flat space-time, the non-relativistic limit $c\rightarrow\infty$ reduces  the Einstein equation to the Poisson  equation which determines the Newtonian potential.
It is known that this space-time is described by the Newton-Cartan geometry.\footnote{For example, see the textbook~\cite{gra} for the Newton-Cartan geometry.}
On the analogy of this case, in the bulk AdS$_{d+1}$ gravity dual to the GCA, one has to  take the scaling limit $t\rightarrow t$, $r\rightarrow r$ and $x^i\rightarrow\epsilon x^i\,\,(i=1,2,\cdots ,d-1)$ with $\epsilon\rightarrow 0$, where $r$ is the radial coordinate in the bulk and can be interpreted as a measure of the energy scale in the boundary theory.
The resulting space-time is a surviving AdS$_2$ base times the degenerate $x^i$ direction, but nevertheless it can be also described by the  Newton-Cartan-like geometry, as discussed in ref.~\cite{gopa1}.
On the basis of this proposal, we want to realize the GCA in two dimensions on the degenerate AdS$_3$ gravity under the scaling limit $t\rightarrow t$, $r\rightarrow r$ and $x\rightarrow\epsilon x$.

We have already known that the usual AdS$_3$/CFT$_2$ correspondence is confirmed by  Brown and Henneaux's analysis~\cite{brown}.
Namely, the  Virasoro algebra is obtained from the algebra of the Hamiltonian with respect to the allowed deformation around the asymptotic infinity $r\rightarrow\infty$ in three dimensional gravity with the negative cosmological constant $-2/\ell^2$~\cite{DJ}.
Its central charges are then expressed by $c^\pm=\frac{3\ell}{2G}$, where $G$ is the Newton constant in three dimensions.
This setup of the left-right symmetric central charges 
for the parent CFT  is not enough to realize the full
  GCA in two dimensions 
even if we take the above corresponding limit  on both boundary and bulk sides.

In order to construct the asymmetry of the central charges, we have to begin with the three dimensional gravity including the gravitational Chern-Simons term, called  topologically massive gravity~\cite{Deser:1982vy}.
We add to it the negative cosmological constant and refer to this three dimensional gravity as cosmological topologically massive gravity (CTMG).\footnote{See ref.~\cite{krauslarsen}-\cite{sachs} for the recent discussions on three dimensional gravity with the gravitational Chern-Simons term.}
By the direct Brown-Henneaux approach to CTMG, it was shown in ref.~\cite{HHKT} that the Virasoro algebra is again observed on the AdS$_3$ boundary and that the central charges then become asymmetric, i.e., $c^\pm=\frac{3\ell}{2G}\left(1\pm\frac{1}{\mu\ell}\right)$, where $1/\mu$ is the coupling of the gravitational Chern-Simons term.
Since the entropy for the Ba\~{n}ados-Teitelboim-Zanelli (BTZ) black hole~\cite{btz} in CTMG again agrees with the Cardy formula~\cite{cardy}, one can confirm the AdS$_3$/CFT$_2$ correspondence even in this higher derivative gravity.\footnote{The equivalent expressions of $c^\pm$ or the entropy in CTMG were obtained in various ways~\cite{krauslarsen}-\cite{guptasen},~\cite{bla},~\cite{HHKNT2}.}

We now argue that 
this asymmetry of $c^\pm$ is compatible with the GCA features, and in fact the additional limit $\mu\rightarrow\epsilon\mu$ gives rise to the central charges $C_1$ and $C_2$. 
By taking  appropriate limits on both sides between the AdS$_3$ gravity in CTMG and the Virasoro algebra for the parent relativistic CFT$_2$, we verify the correspondence between the Newton-Cartan-like geometry in the bulk CTMG and the GCA in two dimensions on the $r\rightarrow \infty$ boundary.
Our proposal is  that the scaling limit $\mu\rightarrow \epsilon\mu$ together with $t\rightarrow t$, $r\rightarrow r$ and $x\rightarrow\epsilon x$ leads to a consistent realization of the GCA with the non-trivial  $C_1$, $C_2$, $\Delta$ and $\xi$.
These parameters are again  expressed by those of gravity.

Now that we have 
 the realization of the GCA in the bulk  CTMG, we can present a new entropy formula of the GCA from the bulk black hole entropy.
Under our limit, the entropy for the BTZ black hole is still finite and can be rewritten by the parameters of the GCA.
Our new entropy formula is
\begin{align}
\mathcal{S}_{\text{GCA}}=\pi\left(C_1\sqrt{\frac{2\xi}{C_2}}+\Delta\sqrt{\frac{2C_2}{\xi}}\right).
\end{align}
We also discuss whether it can be reproduced  by counting the degeneracy of microstates of the GCA in two dimensions.
Then we have to use the modular transformation, which is now  degenerated by  the scaling limit, even after the reduction to the non-relativistic geometry.
This entropy expression is expected to give quantum  degeneracy for the GCA states with the scaling dimensions $\Delta$ and the rapidity $\xi$, and is an analog of the Cardy formula for the relativistic CFT$_2$.

This paper is organized as follows: in section~\ref{GCA2} we first review 
 the infinite dimensional GCA in two dimensions and how it can be reduced from the Virasoro algebra.
In section~\ref{CTMG3} we recall the usual AdS$_3$/CFT$_2$ correspondence in CTMG and show that the scaling limit surely yields the GCA and its finite parameters.
In section~\ref{entropy} we present the new entropy formula from the BTZ black hole entropy in the bulk and try to demonstrate the microscopic counting of Galilean field theory in order to reconstruct  it.
In section~\ref{sum} we summarize our results.


\section{Galilean Conformal Algebra in Two Dimensions}
\label{GCA2}

In two dimensional space-time $(t,x)$, the generators
\begin{align}
\mathcal{L}^\pm_n=ie^{inx^\pm}\partial_\pm,
\end{align}
where $x^\pm=t\pm x$ and  $\partial_\pm=\frac{1}{2}(\partial_t\pm \partial_x)$, obey the (center-less) Virasoro algebra
\begin{align}
[\mathcal{L}^\pm_m,\mathcal{L}^\pm_n]&=(m-n)\mathcal{L}^\pm_{m+n},\notag\\
[\mathcal{L}^+_m,\mathcal{L}^-_n]&=0.
\end{align}
From this, one obtains the infinite dimensional (center-less) GCA in two dimensions  
\begin{align}
[L_m,L_n]&=(m-n)L_{m+n},\notag\\
[L_m,M_n]&=(m-n) M_{m+n},\notag\\
[M_m,M_n]&=0,
\label{GCAc=0}
\end{align}
by taking a non-relativistic limit
\begin{align} 
t\rightarrow t,\,\,\,\,\,\,\,\,
x\rightarrow\epsilon x,\,\,\,\,\,\,\,\,
\textrm{with}\,\,\,\,\,\,\,\,
\epsilon\rightarrow0.
\label{limit}
\end{align}
The GCA generators $L_n$ and $M_n$ are constructed from the Virasoro generators $\mathcal{L}^+_n$ and $\mathcal{L}^-_n$ by
\begin{align}
L_n&=\lim_{\epsilon\rightarrow 0}(\mathcal{L}^+_n+\mathcal{L}^-_n),\notag\\
M_n&=\lim_{\epsilon\rightarrow 0}\epsilon(\mathcal{L}^+_n-\mathcal{L}^-_n),
\label{LM}
\end{align}
or their explicit dependence on the coordinates is
\begin{align}
L_n&=ie^{int}\left(\partial_t+inx\partial_x\right),\notag\\
M_n&=ie^{int}\partial_x.
\end{align}

At the quantum level the GCA   is centrally extended
\begin{align}
[L_m,L_n]&=(m-n)L_{m+n}+C_1m(m^2-1)\delta_{m+n,0},\notag\\
[L_m,M_n]&=(m-n) M_{m+n}+C_2m(m^2-1)\delta_{m+n,0},\notag\\
[M_m,M_n]&=0,
\label{GCA}
\end{align}
for the central charges $C_1$ and $C_2$. 
This is in fact obtained from the Virasoro algebra for the relativistic CFT$_2$ 
\begin{align}
[\mathcal{L}^\pm_m,\mathcal{L}^\pm_n]&=(m-n)\mathcal{L}^\pm_{m+n}+\frac{c^\pm}{12}m(m^2-1)\delta_{m+n,0},\notag\\
[\mathcal{L}^+_m,\mathcal{L}^-_n]&=0,
\label{virasoro}
\end{align}
by the non-relativistic  limit (\ref{limit}).
Accordingly, the central charges $C_1$ and $C_2$ are related to $c^\pm$  of the Virasoro algebra by
\begin{align}
C_1&=\lim_{\epsilon\rightarrow 0}\frac{c^++c^-}{12},\notag\\
C_2&=\lim_{\epsilon\rightarrow 0}\left(\epsilon\frac{c^+-c^-}{12}\right).
\label{c1c2}
\end{align}
Eq. (\ref{c1c2}) tells us that $c^++c^-$ is of order $\mathcal{O}(1)$ while  $c^+-c^-$ must be  $\mathcal{O}(1/\epsilon)$ for the non-trivial $C_1$ and $C_2$.
Namely, $c^+$ and $c^-$ should behave as $\mathcal{O}(1)\pm\mathcal{O}(1/\epsilon)$.\footnote{The modular invariance of the CFT$_2$ requires that $c^+-c^-=0\,\,\textrm{mod}\,\,24$. We implicitly impose this condition.}

Similarly, scaling dimensions $\Delta$ and  rapidity $\xi$,
which are the eigenvalues of $L_0$ and  $M_0$ respectively,  are made by
\begin{align}
\Delta&=\lim_{\epsilon\rightarrow 0}(h^++h^-),\notag\\
\xi&=\lim_{\epsilon\rightarrow 0}\epsilon(h^+-h^-),
\end{align}
where $h^+$ and $h^-$ are the eigenvalues of $\mathcal{L}^+_0$ and $\mathcal{L}^-_0$.
In other words, they have to come from the conformal weight $h^\pm=\mathcal{O}(1)\pm\mathcal{O}(1/\epsilon)$. 
From  the behaviours of the central charges and conformal weights in the 
$\epsilon \to 0$ limit 
we find that  the parent CFT breaks unitarity.
But  this may not cause  a problem for us since such non-unitary CFTs often arise in some cases.  
A primary state $|\Delta,\xi\rangle$, which obeys $L_n|\Delta,\xi\rangle=0$ and $M_n|\Delta,\xi\rangle=0$ $(n>0)$, and its descendants built up by $L_{-n}$ and $M_{-n}$  $(n>0)$  give us a representation of the GCA.
Although the Hilbert space of the GCA is again non-unitary, this nature is inherited from the parent relativistic CFT \cite{gopa2}.


\section{GCA Realization on Cosmological Topologically Massive Gravity}
\label{CTMG3}

We would like to propose a gravity dual of this GCA in two dimensions in the context of the AdS/CFT correspondence.
To realize the properties of the GCA from three dimensional gravity, it is especially notable for us  that the dual theory has to yield the finite $(\Delta,\xi)$ and $(C_1,C_2)$.
To do this we first try to realize the parent CFT$_2$ which has the above  $\epsilon$ dependence. 
Actually, left-right asymmetry between the central charges of the Virasoro algebra is seen in the Brown-Henneaux analysis for the particular  three dimensional gravity, called CTMG. 
The gravity action of CTMG is given by
\begin{align}
\mathcal{I}&=\frac{1}{16\pi G}\int d^3x(\mathcal{L}_{\text{EH}}+\mathcal{L}_{\text{CS}}),\notag\\
\mathcal{L}_{\text{EH}}&=\sqrt{-g}\left(R+\frac{2}{\ell^2}\right),\notag\\
\mathcal{L}_{\text{CS}}&=\frac{ 1 }{2\mu} \sqrt{-g}\,
\epsilon^{\mu\nu\rho}\left(\Gamma^\sigma_{\mu\lambda}\partial_\nu\Gamma^\lambda_{\rho\sigma}+\frac{2}{3}\Gamma^\sigma_{\mu\lambda}\Gamma^\lambda_{\nu\theta}\Gamma^\theta_{\rho\sigma}\right).
\label{CTMG}
\end{align}
Here $\mathcal{L}_{\text{CS}}$ is the gravitational Chern-Simons term and $1/\mu$ is its coupling constant.

The gravitational Chern-Simons term contains the third derivative, but nevertheless, this gravity theory with the higher derivative correction still allows the AdS$_3$ geometry as a solution. 
The vacuum state is the globally AdS$_3$
\begin{align}
ds^2=-\left(1+\frac{r^2}{\ell^2}\right)dt^2+\left(1+\frac{r^2}{\ell^2}\right)^{-1}dr^2+r^2d\phi^2,
\label{vacuum}
\end{align}
and its excited state is the BTZ black hole
\begin{align}
ds^2=-\left(\frac{r^2}{\ell^2}-8Gm+\frac{16G^2j^2}{r^2}\right)dt^2+\frac{dr^2}{\frac{r^2}{\ell^2}-8Gm+\frac{16G^2j^2}{r^2}}
+r^2\left(d\phi+\frac{4Gj}{r^2}dt\right)^2
\label{btz}
\end{align}
with $\phi\sim \phi+2\pi$.
The definitions of mass $M$ and angular momentum $J$ of the BTZ black hole 
are affected by the presence 
of the Chern-Simons term and are related to the
 parameters $m$ and $j$ by
\begin{align}
M=m+\frac{1}{\mu}\frac{j}{\ell^2},\notag\\
J=j+\frac{1}{\mu} m.
\label{mass}
\end{align}
according to refs. \cite{GHHM}-\cite{bouchareb}.
The Bekenstein-Hawking entropy is also obliged to be modified as
\begin{align}
\mathcal{S}_{\textrm{BH}}=\frac{\pi r_+}{2G}+\frac{1}{\mu\ell}\frac{\pi  r_-}{2G}.
\label{waldtachikawa}
\end{align}
Here $r_{+}$ and $r_{-}$ are outer and inner horizon radii  defined by 
\begin{align}
r_\pm&=\sqrt{2G\ell\left(\ell m+j\right)}\pm \sqrt{2G\ell\left(\ell m-j\right)}.\label{eq:innerouterradii}
\end{align}

Since the AdS$_3$ solutions are not changed despite the modification of the equations of motion, one can apply the Brown-Henneaux analysis for the AdS$_3$/CFT$_2$ correspondence.
First, let us allow the deformation of the global AdS$_3$ solution at the spatial infinity $(r\rightarrow\infty)$ such as  $\delta g_{\mu\nu}=\mathcal{D}_\mu\xi_\nu+\mathcal{D}_\nu\xi_\mu$. 
The choice
\begin{align}
\xi_n^\pm=e^{inx^\pm}\left(
\partial_\pm-\frac{n^2\ell^2}{2r^2}\partial_\mp-\frac{inr}{2}\partial_r
\right),
\label{xi}
\end{align}
where
$x^\pm=t/\ell\pm \phi$ and $\partial_\pm=\frac{1}{2}(\ell\partial_t\pm \partial_\phi)$, means that the allowed deformation is
\begin{align}
 &g_{tt}=-\frac{r^2}{\ell ^2}+\mathcal{O}(1)
 \,\,\,\,\,\,,\,\,\,\,\,\,
 g_{tr}=\mathcal{O}(r^{-3})
 \,\,\,\,\,\,,\,\,\,\,\,\,
 g_{t\phi}=\mathcal{O}(1), 
 \notag \\
 &g_{rr}=\frac{\ell ^2}{r^2}+\mathcal{O}(r^{-4})
 \,\,\,\,\,\,,\,\,\,\,\,\,
 g_{r\phi}=\mathcal{O}(r^{-3})
 \,\,\,\,\,\,,\,\,\,\,\,\,
 g_{\phi\phi}=r^2+\mathcal{O}(1). 
\label{hypersurface bd}
\end{align}
The BTZ black hole satisfies this behaviour.

The algebraic structure of symmetric transformation group is given by the Poisson bracket algebra of the Hamiltonian generator $H[\xi]$.
But the {\it asymptotic isometry} such as (\ref{xi}) leads to the infinite 
dimensional Lie algebra with central extension. 
It was shown that the Poisson bracket algebra of  $H[\xi]$ in three dimensional gravity with the negative cosmological constant corresponds to the centrally extended Virasoro algebra (\ref{virasoro}).
This is also true in the case of CTMG (\ref{CTMG}), but the central charges of the Virasoro algebra are left-right asymmetric \cite{HHKT}
\begin{align}
c^+&=\frac{3\ell}{2G}\left(1+\frac{1}{\mu\ell}\right),\notag\\
c^-&=\frac{3\ell}{2G}\left(1-\frac{1}{\mu\ell}\right)
\label{clcr}
\end{align}
due to the coupling of the gravitational Chern-Simons term $1/\mu$.\footnote{In order to guarantee the modular invariance of the CFT$_2$, CTMG is required to hold $3/\mu G=0\,\,\textrm{mod}\,\,24$. Here we treat $\mu$ as a continuous parameter by choosing the Newton constant $G$ small enough.}
For the BTZ black hole solution in CTMG, 
$h^+$ and $h^-$ are calculated as\footnote{In our notation, $h^\pm$ contain the factor $c^\pm/24$. The vacuum (\ref{vacuum}) has $M=-1/8G$ and $J=-1/8G\mu$ relative to the zero-mass BTZ, so $h^\pm$ are normalized so that it vanishes for the vacuum. But as long as we consider large BTZ black holes $(mG\gg 1)$, these terms are neglected.}
\begin{align}
h^+=\frac{1}{2}\left(\ell M+J\right)+\frac{c^+}{24},\,\,\,\,\,\,\,\,\,\,h^-=\frac{1}{2}\left(\ell M-J\right)+\frac{c^-}{24}.
\label{hpm}
\end{align}
Then the microscopic state counting is expressed by the Cardy formula
\begin{align}
\mathcal{S}_{\textrm{CFT}}=2\pi\left(\sqrt{\frac{c^+ h^+}{6}}+\sqrt{\frac{c^- h^-}{6}}\right).
\label{cardy}
\end{align}
Even if we include the higher derivative term $\mathcal{L}_{\textrm{CS}}$, this microscopic entropy  still agrees with the black hole entropy (\ref{waldtachikawa}) in $1\leq |\mu|\ell$ case.

Now let us consider a limit corresponding to (\ref{limit}) in three dimensional gravity.
Here we take the scaling limit
\begin{align}
t\rightarrow t,\,\,\,\,\,\,\,\,\,\,
r\rightarrow r,\,\,\,\,\,\,\,\,\,\,
\phi\rightarrow \epsilon \phi,\,\,\,\,\,\,\,\,\,\,
\textrm{with}\,\,\,\,\,\,\,\,
\epsilon\rightarrow0.
\label{gravitylimit}
\end{align}
Accordingly, the parameters $m$ and $j$ in the BTZ solution (\ref{btz}) must scale like
\begin{align}
m\rightarrow m,\,\,\,\,\,\,\,\,\,\,j\rightarrow \epsilon j.
\label{mjlimit}
\end{align}
The limits (\ref{gravitylimit}) and (\ref{mjlimit})  make the Riemannian metric degenerate in the solution (\ref{btz}), and then the geometry seems to be ill-defined. 
But this is similar to the usual Newtonian approximation $c\rightarrow\infty$. 
In ref.~\cite{gopa1}, it was discussed that this situation in bulk  gravity is described by the Newton-Cartan-like geometry for the geometry with the AdS$_2$ base.
The dynamical variables are affine connections, not metric.
And the Einstein equation\footnote{In the case of CTMG the field equation is affected by the contribution of the gravitational Chern-Simons term, but the correction term vanishes as long as we consider the AdS$_3$ solution.} reduces to equation for the curvature of these non-metric connections, just as it does to the Poisson equation for determining the Newtonian gravitational potential in the non-relativistic limit of the asymptotically flat space-time.  
The difference is only the point that in our case $t$ and $r$ have the same scaling on $\epsilon$.
We have the geometric structure of a fibre bundle with the surviving AdS$_2$ base times the degenerate $\phi$ direction.
In other words, the gravity theory is not ill-defined under the scaling limit but still gives the meaningful physical quantities, for example, the entropy as mentioned below.

Since the Virasoro generators correspond to $\mathcal{L}_n^\pm=i\xi_n^\pm$, from the gravity side we define the GCA generators $L_n$ and $M_n$ as
\begin{align}
L_n&=ie^{int/\ell}\left[\ell\left(1-\frac{n^2\ell^2}{2r^2}\right)\partial_t+in\phi\left(1+\frac{n^2\ell^2}{2r^2}\right)\partial_\phi-inr\partial_r\right],\notag\\
M_n&=ie^{int/\ell}\left(1+\frac{n^2\ell^2}{2r^2}\right)\partial_\phi,
\end{align}
using (\ref{xi}), (\ref{gravitylimit}) and (\ref{LM}).
One can easily check that they certainly satisfy the center-less GCA algebra (\ref{GCAc=0}).
Recalling that the central charges of the Virasoro must behave as $c^\pm=\mathcal{O}(1)\pm\mathcal{O}(1/\epsilon)$, we assume that $\mu$ should scale as 
\begin{align}
\mu\rightarrow \epsilon \mu.
\label{betalimit}
\end{align}
Under this limit, we obtain the central charges of the GCA
\begin{align}
C_1&=\frac{\ell}{4G},\notag\\
C_2&=\frac{1}{4G\mu}\label{GCAc}
\end{align}
from their definitions (\ref{c1c2}) and the Virasoro central charges (\ref{clcr}).
We have not known yet  the corresponding {\it canonical formalism  for the Newton-Cartan-like geometry}, but we have got around it and have obtained 
 the centrally extended GCA on the $r\rightarrow\infty$ boundary 
 by the scaling limit (\ref{gravitylimit}), (\ref{mjlimit}) and (\ref{betalimit}) from the Virasoro algebra of the Brown-Henneaux analysis in AdS$_3$ gravity.
The scaling of $\mu$ is also consistent with the finite $\Delta$ and $\xi$, and they are given by
\begin{align}
\Delta&=\lim_{\epsilon\rightarrow 0}\left(\ell M+\frac{c^++c^-}{24}\right)=\ell m+\frac{1}{\mu\ell}j+\frac{C_1}{2},\notag\\
\xi&=\lim_{\epsilon\rightarrow 0}\epsilon \left(J+\frac{c^+ - c^-}{24}\right)=\frac{1}{\mu} m+\frac{C_2}{2}
\label{delta}
\end{align}
for the parameters of gravity.
When $m$ and $j$ are large enough, the last terms $C_1/2$ and $C_2/2$ can be neglected as usual. 
In conclusion, the limits (\ref{gravitylimit}), (\ref{mjlimit}) 
and (\ref{betalimit}) define the consistent GCA on 
the $r\rightarrow\infty$ boundary of the bulk  gravity.
Eq. (\ref{mass}) tells us that the mass and angular momentum of the 
BTZ black hole behave in the scaling limit (\ref{mjlimit}) and (\ref{betalimit}) as
\begin{align}
M \to M, 
\hskip1cm
J \sim {\cal O}\left (\frac{1}{\epsilon}\right ). \label{eq29}
\end{align}
The scaling $j\to \epsilon j$ in (\ref{mjlimit}) has been motivated so that the black hole mass $M$ should remain finite under the scaling.
Note that 
the behavior of $J$ in (\ref{eq29}) is consistent with (\ref{gravitylimit}) since 
$J \sim \frac{\displaystyle{\partial}}{\displaystyle{\partial \phi}}
\to  \frac{\displaystyle{1}}{\displaystyle{\epsilon}}
\frac{\displaystyle{\partial} }{\displaystyle{\partial \phi}}$.


\section{Entropy Formula of GCA}
\label{entropy}

Finally, we would like to mention the entropy of the GCA. 
The scaling limit (\ref{mjlimit}) requires that  the event horizon of the BTZ black hole 
(\ref{eq:innerouterradii})
should scale as
\begin{align}
r_+\rightarrow\,2\ell\sqrt{2Gm},
\,\,\,\,\,\,\,\,\,\,\,\,\,\,\,
r_-\rightarrow\,\epsilon\sqrt{\frac{2G}{m}}j.
\end{align}
Since the radius of the inner horizon behaves as $r_-\sim \mathcal{O}(\epsilon)$ while the scaling of $\mu$ is (\ref{betalimit}), the black hole entropy (\ref{waldtachikawa}) is still finite
\begin{align}
\mathcal{S}_{\textrm{BH}}\rightarrow \pi\left(\ell\sqrt{\frac{2m}{G}}+\frac{ j}{\mu\ell}\sqrt{\frac{1}{2Gm}}\right).
\end{align}
Here we have implicitly used the fact that the periodicity of $\phi$ now becomes $\phi\sim\phi+\frac{2\pi}{\epsilon}$.
From the expressions of the central charges $(C_1,C_2)$,  the scaling dimensions $\Delta$ and the rapidity $\xi$, one can rewrite the entropy as
\begin{align}
\mathcal{S}_{\text{GCA}}=\pi\left(C_1\sqrt{\frac{2\xi}{C_2}}+\Delta\sqrt{\frac{2C_2}{\xi}}\right)
\label{GCAS}
\end{align}
for  $m \gg 1/G$ and $j \gg \mu \ell ^{2}/G$.
We here propose that this expression   is  the entropy for the GCA in two dimensions, which is an analog of the Cardy formula for the relativistic CFT$_2$.
Eq. (\ref{GCAS}) is the most important formula in the present work.

There are some comments on the entropy formula (\ref{GCAS}).
The black hole entropy (\ref{waldtachikawa}) is derived in the Noether charge formulation, and one can show that it obeys the first law of thermodynamics~\cite{wald,jkm,iyer,tachikawawald}.
However, it  is in general not trivial to prove the second law in the higher derivative gravity like CTMG, except for the Bekenstein-Hawking piece,  the first 
term in (\ref{waldtachikawa}).
Furthermore,  when we naively take $1\gg\mu\ell$,  
the correction piece to the Bekenstein-Hawking, 
for which the second law is not manifestly  guaranteed,  becomes dominant.\footnote{
In such cases, if we have 
only the first law at our hands, we may consider the {\it exotic black holes}, 
for  which the roles of the outer and inner horizon  are interchanged, with the entropy
\begin{align}
\mathcal{S}_{\textrm{ex}}=\frac{\pi r_-}{2G}+\frac{1}{\mu\ell}\frac{\pi  r_+}{2G}.\notag
\end{align}
If we could evaluate the first law at the inner horizon, the exotic black hole entropy  would satisfy the re-defined first law~\cite{exotic,park}.
}
However, as long as we take the limit (\ref{mjlimit}) and (\ref{betalimit}) at the same time instead of the naive limit $1\gg\mu\ell$,  both the first and second 
terms of the entropy (\ref{waldtachikawa}) are of the same order, as seen above.
It is an important task for us to confirm the second law, but we here adopted the usual entropy formula derived by the Noether charge method.

It is also notable that the form (\ref{cardy}) would become divergent under the $\epsilon\rightarrow0$ limit, unlike the black hole entropy (\ref{waldtachikawa}).
In $1>\mu\ell$ region both $c^-$ and $h^-$ are negative, so we might consider the new representation of the Virasoro algebra with $\tilde{\mathcal{L}}^-_n\equiv-\mathcal{L}_{-n}^-$, $\tilde{c}^-\equiv-c^-$ and the highest-weight state $|\tilde{h}^-\rangle$ satisfying $\tilde{\mathcal{L}}^-_n|\tilde{h}^-\rangle=0$ $(n>0)$.
Counting the relativistic primary states $|h^+,\tilde{h}^-\rangle$ and their 
 descendants, one again arrives at the expression (\ref{cardy}), and it becomes divergent in $\epsilon\rightarrow 0$.   
However, we alternatively know the  representation of the GCA in our limit, and now  want to count the degeneracy of the GCA primary states 
$|\Delta,\xi\rangle$ and their descendants.
Then the entropy actually becomes finite, although they  include the 
negative norm  states.
But remarkably, the GCA states $|\Delta,\xi\rangle$ are completely different from  the states $\lim_{\epsilon\rightarrow 0}|h^+,\tilde{h}^-\rangle$ since they come from $|h^+,h^-\rangle$ satisfying $\mathcal{L}^\pm_n|h^+,h^-\rangle=0$ $(n>0)$, not $|h^+,\tilde{h}^-\rangle$.
It is, therefore, natural that the expression (\ref{cardy}) does not give the same answer as ours (\ref{GCAS}).
Here we discuss whether the direct state counting of the GCA representation can confirm the entropy formula (\ref{GCAS}).

From now on, we take the Euclidean signature for convenience.
The partition function for the relativistic CFT$_2$ on the cylinder is written 
in  the usual form
\begin{align}
Z_0(\beta,\theta)=\textrm{Tr}\left[e^{-\beta\left(\mathcal{L}_0^++\mathcal{L}_0^--\frac{c^+}{24}-\frac{c^-}{24}\right)}e^{i\theta\left(\mathcal{L}_0^+-\mathcal{L}_0^--\frac{c^+}{24}+\frac{c^-}{24}\right)}\right],
\end{align}
where $\beta$ is the inverse temperature and $\theta$ the ``angular potential".
We here want to consider the non-relativistic limit of this parent CFT$_2$.
Then, with the help of eqs. (\ref{limit}), (\ref{LM}) and (\ref{c1c2}), it is 
reduced to the form
\begin{align}
Z_0(\beta,\theta)=\textrm{Tr}\left[e^{-\beta \left(L_0-\frac{C_1}{2}\right)+i\theta \left(M_0-\frac{C_2}{2}\right)}\right].
\end{align}
When the partition function on the  torus is evaluated as
\begin{align}
Z(\beta,\theta)&=e^{-\beta\frac{ C_1}{2}+i\theta \frac{C_2}{2}}Z_0(\beta,\theta)\notag\\
&=\int^\infty_{\Delta_0}d\Delta \,e^{-\beta\Delta}\int^\infty_{\xi_0}d\xi \,e^{i\theta\xi}\,\Omega(\Delta,\xi),
\label{torus}
\end{align}
the inverse Laplace transformation gives us 
\begin{align}
\Omega(\Delta,\xi)=\frac{1}{(2\pi)^2 i}\int^{\beta'+i\infty}_{\beta'-i\infty} d\beta \,e^{\beta\Delta-\beta\frac{C_1}{2}}\int^{\infty}_{-\infty}d\theta\, e^{-i\theta\xi+i\theta\frac{C_2}{2}}\,Z_0(\beta,\theta).
\end{align}
This quantity is an analog of the the degeneracy for the state 
with $(\Delta,\xi)$, but further details on microscopic interpretation 
will be discussed later.

To derive the Cardy formula (\ref{cardy}), one had to use the modular invariance of the torus for the CFT$_2$.
However, the non-relativistic limit is now taken on the two dimensional space-time.
The modular transformation  for the torus is translated  as 
\begin{align}
\beta
\leftrightarrow 
\frac{4\pi^2}{\beta},\,\,\,\,\,\,\,\,\,\,
\theta
\leftrightarrow-\frac{4\pi^2}{\beta^2}\theta
\label{modular}
\end{align}
in our limit.
The invariance of the partition function 
under  this transformation has to be tested from various viewpoints, 
but  now we  simply assume this remnant of the modular invariance for the parent CFT.   
Imposing this invariance of $Z_0(\beta,\theta)$, which amounts to  the relation
\begin{align}
Z_0(\beta,\theta)=Z_0(4\pi^2/\beta,-4\pi^2\theta/\beta^2)=e^{\frac{2\pi^2 C_1}{\beta}+i\theta\frac{2\pi^2 C_2}{\beta^2}}
Z(4\pi^2/\beta,-4\pi^2\theta/\beta^2),
\end{align}
then one finds
\begin{align}
\Omega(\Delta,\xi)=\frac{1}{(2\pi )^2i}\int d\beta \int d\theta\, e^{-i\theta\left(\xi-\frac{C_2}{2}-\frac{2\pi^2C_2}{\beta^2}\right)}e^{-\frac{\beta C_1}{2}+\frac{2\pi^2 C_1}{\beta}+\beta\Delta}\,Z(4\pi^2/\beta,-4\pi^2\theta/\beta^2).
\label{eq:degeneracy}
\end{align}
Without detailed knowledge on the partition function, we can
approximately evaluate this integral for large charges $\Delta$ and $\xi$
 (equivalently for large $m$ and $j$ on the gravity side).
By performing the $\theta$-integration first, we obtain a sharp Gaussian contribution around the point
\begin{align}
\beta = \sqrt{\frac{2\pi^2C_2}{\xi-C_2/2}} \simeq \sqrt{\frac{2\pi^2C_2}{\xi}},
\label{eq:saddlepoint}
\end{align}
where  $Z(4\pi ^{2}/\beta , -4\pi ^{2}\theta /\beta ^{2})$ varies slowly as a function of $\theta $. 
 If we move the path of $\beta$-integration in advance so that it passes through this point (\ref{eq:saddlepoint}), the integral over $\beta$ is dominated by the contribution from  (\ref{eq:saddlepoint}). Then the logarithm of 
 (\ref{eq:degeneracy})  can be approximately written as
\begin{align}
 \log \Omega(\Delta, \xi) &\simeq  2\pi^2C_1\sqrt{\frac{\xi}{2\pi^2C_2}} + \Delta\sqrt{\frac{2\pi^2C_2}{\xi}}
\nonumber \\
 &= \pi\left(C_1\sqrt{\frac{2\xi}{C_2}} + \Delta\sqrt{\frac{2C_2}{\xi}}\right),
\end{align}
which coincides with the entropy formula (\ref{GCAS}) proposed from the gravity-side analysis. This is the non-relativistic counterpart of the Cardy formula.



\section{Summary and Discussions}
\label{sum}

In this paper, we have considered  the realization of the infinite dimensional GCA in two dimensions on the boundary of the three dimensional gravity, CTMG.
The scaling limit on the AdS$_3$/CFT$_2$ correspondence by Brown-Henneaux's method enables us to confirm the consistent correspondence between the Newton-Cartan-like geometry with AdS$_2$ base and the GCA with the non-zero centers.
The compatible limit on the bulk side  is given by (\ref{gravitylimit}), (\ref{mjlimit}) and (\ref{betalimit}).
The central charges $C_1$ and $C_2$, the scaling dimensions $\Delta$ and the rapidity $\xi$ are expressed as   (\ref{GCAc}) and (\ref{delta}) in terms of the parameters of gravity.
Moreover,  we have proposed the entropy formula of the GCA (\ref{GCAS}) from both the bulk and boundary analyses. 
To justify this, we have used the transformation (\ref{modular}) inherited from the modular invariance of the parent CFT.
The meaning of this transformation should be clarified  further without 
referring to the parent CFT.

We  give some additional comments on the non-relativistic analog of the Cardy formula (\ref{GCAS}). Since there are states with negative norms in the 
Galilean field theory, one might think that $\Omega(\Delta, \xi)$ is the number of positive norm states minus that of negative norm states with the given charges $\Delta$ and $\xi$. But the situation is more complicated. In the two dimensional Galilean field theory, the rapidity operator $M_0$ is non-diagonalizable, because the representation matrix of $M_0$ has a Jordan cell structure when it is restricted to the subspace spanned by the GCA descendants with the same scaling dimensions in a GCA module~\cite{gopa2}. This means that the partition function $Z(\beta,\theta)$ cannot be written as a sum of the (plus or minus of) Boltzmann factors for all the states, such as
\begin{align}
 \sum_{n\in{\rm states}} (\pm ) e^{-\beta \Delta_n +i\theta \xi_n}.
\end{align}
It actually has  logarithmic contributions like\footnote{For example, suppose the subspace spanned by the level-one descendants $L_{-1}\left|\Delta-1, \xi\right>$ and $M_{-1}\left|\Delta-1, \xi\right>$, where $\left|\Delta-1, \xi\right>$ is a GCA primary that has the scaling dimensions $\Delta-1$ and the rapidity $\xi$. The representation matrix of $e^{i\theta M_0}$ restricted to this subspace can be written as $\left(\begin{array}{cc}e^{i\theta\xi} & i\theta e^{i\theta\xi}\\ 0 & e^{i\theta\xi}\\\end{array}\right)$. Since $\left<\Delta-1,\xi\right|L_{1}M_{-1}\left|\Delta-1,\xi\right>\neq 0$, the trace of $e^{-\beta L_0 + i\theta M_0}$ over this subspace leads to the logarithmic contribution such as (\ref{log}).}
\begin{align}
 i\theta\, e^{-\beta \Delta + i\theta \xi}.\label{log}
\end{align}
This situation is rather similar to  the case of the logarithmic CFT~\cite{gurarie}. It has the Jordan cell structure with respect to $\mathcal{L}_0$, and the correlation functions and the partition function have the logarithmic contributions. Nevertheless, modular invariant partition functions of logarithmic CFTs are known to exist~\cite{flohr}. 
Considering the existence of the logarithmic terms (\ref{log}), we have to 
learn more about the interpretation of  the inverse Laplace transform 
$\Omega(\Delta, \xi)$ as the degeneracy of  the states. 
We would like to emphasize, however,  that $\log \Omega (\Delta, \xi)$ certainly agrees with the black hole entropy (\ref{GCAS}) for large $\Delta $ and $\xi$. 

Also,  the correlation functions of the GCA are calculated in ref.~\cite{gopa2}, and they are consistent with those obtained from  the Virasoro algebra by taking the corresponding limit.
As a next step for  strengthening  the relation between GCA and CTMG, it is a non-trivial check whether they can be re-derived by the holographic principle based on the GKPW relation.
In refs.~\cite{marika,sachs} the correlation functions for the logarithmic CFT were computed in the AdS$_3$ bulk of CTMG at the chiral point~\cite{lss}.
It is a quite intriguing and significant problem for the future works  to investigate the connection between the GCA and the logarithmic CFT.\footnote{The logarithmic correlators in the two dimensional Galilean field theory were argued in ref.~\cite{hosseiny-rouhani2}. }


\section*{Acknowledgements}

We wish to thank   Yoshifumi Hyakutake,  Hiroaki Tanida and Satoshi Yamaguchi for many useful discussions.
K.H. and T.N. are supported in part by JSPS Research Fellowship for Young Scientists.






\begin{thebibliography}{99} 



\bibitem{nishida}
Y. Nishida and D.T. Son,
``{\em
Nonrelativistic conformal field theories,}''
Phys. Rev. D {\bf76} (2007) 086004, arXiv:0706.3746 [hep-th].

\bibitem{son}
D.T. Son, 
``{\em
 Toward an AdS/cold atoms correspondence: A Geometric realization of the Schrodinger symmetry,}''
Phys. Rev. D {\bf78} (2008) 046003, arXiv:0804.3972 [hep-th].

\bibitem{bala}
K. Balasubramanian and J. McGreevy, 
``{\em
 Gravity duals for non-relativistic CFTs,}''
Phys. Rev. Lett. {\bf101} (2008) 061601, arXiv:0804.4053 [hep-th].


\bibitem{gopa1}
A. Bagchi and R. Gopakumar,
``{\em 
 Galilean Conformal Algebras and AdS/CFT,}''
JHEP {\bf 0907} (2009) 037, arXiv:0902.1385 [hep-th].

\bibitem{}
M. Alishahiha, A. Davody and A. Vahedi,
``{\em
On AdS/CFT of Galilean Conformal Field Theories,}''
JHEP {\bf0908} (2009) 022, arXiv:0903.3953 [hep-th]. 

\bibitem{bagchi}
A. Bagchi and I. Mandal,
``{\em
On Representations and Correlation Functions of Galilean Conformal Algebras,}''
Phys. Lett. B {\bf 675} (2009) 393, arXiv:0903.4524 [hep-th].

\bibitem{}
D. Martelli and Y. Tachikawa, ``{\em
 Comments on Galilean conformal field theories and their geometric realization,}'' arXiv:0903.5184 [hep-th]. 

\bibitem{}
C. Duval and  P.A. Horvathy,
``{\em
Non-relativistic conformal symmetries and Newton-Cartan structures,}''
J. Phys. A {\bf42} (2009) 465206, arXiv:0904.0531 [math-ph].

\bibitem{}
J.A. de Azcarraga and J. Lukierski,
``{\em
Galilean Superconformal Symmetries,}''
 Phys. Lett. B {\bf 678} (2009) 411, arXiv:0905.0141 [math-ph].

\bibitem{}
M. Sakaguchi,
``{\em
 Super Galilean conformal algebra in AdS/CFT,}'' arXiv:0905.0188 [hep-th]. 

\bibitem{}
A. Bagchi and I. Mandal,
``{\em
Supersymmetric Extension of Galilean Conformal Algebras,}''
Phys. Rev. D {\bf 80} (2009) 086011, arXiv:0905.0580 [hep-th].

\bibitem{}
A. Mukhopadhyay,
``{\em
A Covariant Form of the Navier-Stokes Equation for the Galilean Conformal Algebra,}''
JHEP {\bf 1001} (2010) 100, arXiv:0908.0797 [hep-th]. 

\bibitem{hosseiny-rouhani}
A. Hosseiny and S. Rouhani,
``{\em
Affine Extension of Galilean Conformal Algebra in 2+1 Dimensions,}'' arXiv:0909.1203 [hep-th]. 

\bibitem{gopa2}
A. Bagchi, R. Gopakumar, I. Mandal and A. Miwa,
``{\em GCA in 2d,}''
arXiv:0912.1090 [hep-th]. 

\bibitem{hosseiny-rouhani2}
A. Hosseiny and S. Rouhani,
``{\em Logarithmic Correlators in Non-relativistic Conformal Field Theory,}''
arXiv:1001.1036 [hep-th].

\bibitem{man}
I. Mandal,
``{\em
 Supersymmetric Extension of GCA in 2d,}'' arXiv:1003.0209 [hep-th].


\bibitem{gra}
C.W. Misner,  K.S. Thorne and J.A. Wheeler,  
``{\em
Gravitation,}''
San Francisco 1973, 1279p.

\bibitem{brown}
J.D. Brown and M. Henneaux, ``{\em  Central Charges in the Canonical Realization of Asymptotic Symmetries: An Example from Three-Dimensional Gravity,}''
Commun. Math. Phys. {\bf 104} (1986) 207.




\bibitem{DJ}
S. Deser and R. Jackiw,
``{\em  Three-Dimensional Cosmological Gravity: Dynamics of Constant Curvature,}''
 Ann. Phys. {\bf 153} (1984) 405.

\bibitem{Deser:1982vy}
  S. Deser, R. Jackiw and S. Templeton,
  ``{\em Three-Dimensional Massive Gauge Theories,}''
  Phys. Rev. Lett.  {\bf 48} (1982) 975;
  ``{\em Topologically Massive Gauge Theories,}''
  Ann. Phys.\  {\bf 140} (1982) 372
  [Erratum-ibid.\  {\bf 185} (1988) 406];
  Ann. Phys.\  {\bf 281} (2000) 409.















\bibitem{krauslarsen}
P. Kraus and F. Larsen,
``{\em  Microscopic Black Hole Entropy in Theories with Higher Derivatives,}''
JHEP {\bf 0509}  (2005) 034, hep-th/0506176;
``{\em  Holographic Gravitational Anomalies,}''
JHEP {\bf 0601}  (2006) 022, hep-th/0508218.

\bibitem{solodukhin}
S.N. Solodukhin,
``{\em  Holography with Gravitational Chern-Simons,}''
Phys. Rev. D {\bf 74}  (2006) 024015, hep-th/0509148.


\bibitem{sahoo}
B. Sahoo and A. Sen, ``{\em BTZ Black Hole with Chern-Simons and Higher Derivative Terms,}''
JHEP {\bf 0607} (2006) 008, hep-th/0601228.





\bibitem{park}
M.-I. Park,
``{\em  BTZ Black Hole with Gravitational Chern-Simons: Thermodynamics and Statistical Entropy,}''
Phys. Rev. D {\bf  77} (2008) 026011,  hep-th/0608165;
``{\em  BTZ Black Hole with Higher Derivatives, the Second Law of Thermodynamics, and Statistical Entropy: A New Proposal,}''
Phys. Rev. D {\bf  77} (2008) 126012,  hep-th/0609027.

\bibitem{tachikawawald}
Y. Tachikawa,
``{\em  Black Hole Entropy in the Presence of Chern-Simons Terms,}''
Class. Quant. Grav. {\bf 24} (2007) 737,  hep-th/0611141.

\bibitem{park2}
M.-I. Park,
``{\em   Holography in Three-dimensional Kerr-de Sitter Space with a Gravitational Chern-Simons Term,}''
Class. Quant. Grav. {\bf  25} (2008) 135003,  arXiv:0705.4381 [hep-th]. 




\bibitem{guptasen}
R.K. Gupta and A. Sen,
``{\em  Consistent Truncation to Three Dimensional (Super-) gravity,}''
JHEP {\bf  0803} (2008) 015,  arXiv:0710.4177 [hep-th].


\bibitem{lss}
W. Li, W. Song and  A. Strominger, ``{\em Chiral Gravity in Three Dimensions,}''
JHEP {\bf 0804} (2008) 082, arXiv:0801.4566[hep-th].



\bibitem{cdww}
S. Carlip, S. Deser, A. Waldron and D.K. Wise,
``{\em  Cosmological Topologically Massive Gravitons and Photons,}'' Class. Quant. Grav. {\bf  26} (2009) 075008, arXiv:0803.3998 [hep-th];
``{\em
Topologically Massive AdS Gravity,}''
Phys. Lett. B {\bf666} (2008) 272, arXiv:0807.0486 [hep-th].

\bibitem{HHKT} 
K. Hotta, Y. Hyakutake, T. Kubota, and H. Tanida,
	``{\em Brown-Henneaux's Canonical Approach to Topologically
	Massive Gravity},'' JHEP {\bf 0807} (2008)  066, arXiv:0805.2005 [hep-th].


%
\bibitem{grumiller}
D. Grumiller and N. Johansson, ``{\em  Instability in Cosmological Topologically Massive Gravity at the Chiral Point,}'' JHEP {\bf 0807} (2008) 134, arXiv:0805.2610 [hep-th].
%

\bibitem{}
S. Carlip, ``{\em
 The Constraint Algebra of Topologically Massive AdS Gravity,}''
JHEP {\bf0810} (2008) 078, arXiv:0807.4152 [hep-th];
``{\em
Chiral Topologically Massive Gravity and Extremal B-F Scalars,}'' JHEP {\bf0909} (2009) 083, arXiv:0906.2384 [hep-th].

\bibitem{giribet}
G. Giribet, M. Kleban and M. Porrati, ``{\em  Topologically Massive Gravity at the Chiral Point is Not Chiral,}'' JHEP {\bf 0810} (2008) 045, arXiv:0807.4703 [hep-th].

\bibitem{myung}
Y.-S. Myung, ``{\em Logarithmic Conformal Field Theory Approach to Topologically Massive Gravity,}'' Phys. Lett. B {\bf 670} (2008) 220, arXiv:0808.1942[hep-th].

\bibitem{}
M. Henneaux, C. Martinez and R. Troncoso, 
``{\em
Asymptotically anti-de Sitter spacetimes in topologically massive gravity,}''
 Phys. Rev. D {\bf79} (2009) 081502, arXiv:0901.2874 [hep-th].




%
\bibitem{mss}
A. Maloney, W. Song and A. Strominger,
``{\em  Chiral Gravity, Log Gravity and Extremal CFT,}'' arXiv:0903.4573 [hep-th].

\bibitem{HHKNT2}
K. Hotta, Y. Hyakutake, T. Kubota, T. Nishinaka and H. Tanida, 
``{\em Left-Right Asymmetric Holographic RG Flow with Gravitational Chern-Simons Term,}''
Phys. Lett. B {\bf 680} (2009) 279, arXiv:0906.1255 [hep-th].

\bibitem{marika}
K. Skenderis, M. Taylor and B.C. van Rees, 
``{\em Topologically Massive Gravity and the AdS/CFT Correspondence,}''
JHEP {\bf0909} (2009) 045, arXiv:0906.4926 [hep-th].

\bibitem{}
O. Miskovic and R. Olea, 
``{\em Background-independent charges in Topologically Massive Gravity,}''
JHEP {\bf 0912} (2009) 046, arXiv:0909.2275 [hep-th]. 

\bibitem{sachs}
D. Grumiller and I. Sachs, ``{\em
AdS3/LCFT2 -- Correlators in Cosmological Topologically Massive Gravity,}'' arXiv:0910.5241 [hep-th].

\bibitem{btz}
M. Ba${\tilde {\rm n}}$ados, C. Teitelboim and J. Zanelli, ``{\em The Black Hole in Three-dimensional Space-time,}''
Phys. Rev. Lett. {\bf 69} (1992) 1849, hep-th/9204099. 

\bibitem{cardy}
J.L. Cardy, ``{\em
Operator Content of Two-Dimensional Conformally Invariant Theories,}''
Nucl. Phys. B {\bf 270} (1986) 186.

\bibitem{bla}
M. Blagojevic and B. Cvetkovic, ``{\em
 Canonical structure of 3-D gravity with torsion,}''
'Trends in General Relativty and Quantum Cosmology'. Volume 2. Edited by Charles V. Benton. N.Y., Nova Science Publishers, 2006. pp.85 gr-qc/0412134.




 
\bibitem{GHHM} 
A.A. Garcia, F.W. Hehl, C. Heinicke and A. Macias,
``{\em
Exact Vacuum Solution of a (1+2)-dimensional Poincare Gauge Theory: BTZ Solution with Torsion,}''
 Phys. Rev. D {\bf 67} (2003) 124016, gr-qc/0302097.

\bibitem{MCL} 
K.A. Moussa, G. Clement and C. Leygnac,
``{\em
The Black Holes of Topologically Massive Gravity,}''
 Class. Quant. Grav. {\bf 20} (2003) L277, gr-qc/0303042.


\bibitem{DT}
S. Deser and B. Tekin,
``{\em
 Energy in Topologically Massive Gravity,}''
 Class. Quant. Grav. {\bf 20} (2003) L259, gr-qc/0307073.

\bibitem{DKT}
S. Deser, I. Kanik and B. Tekin,
``{\em Conserved Charges of Higher D Kerr-AdS Spacetimes,}'' Class. Quant. Grav. {\bf 22} (2005) 3383, gr-qc/0506057.

\bibitem{OST}
S. Olmez, O. Sarioglu and B. Tekin, ``{\em Mass and Angular Momentum of Asymptotically AdS or Flat Solutions in the Topologically Massive Gravity,}'' 
Class. Quant. Grav. {\bf 22} (2005) 4355, gr-qc/0507003.

\bibitem{bouchareb}
A. Bouchareb and G. Clement,
``{\em Black Hole Mass and Angular Momentum in Topologically Massive Gravity,}''
Class. Quant. Grav. {\bf 24} (2007) 5581, arXiv:0706.0263 [gr-qc].




\bibitem{wald}
R.M. Wald, ``{\em Black Hole Entropy is the Noether Charge,}''
Phys. Rev. D {\bf 48} (1993) R3427, gr-qc/9307038.

\bibitem{jkm}
T. Jacobson, G. Kang and R.C. Myers, 
``{\em
On Black Hole Entropy,}''
Phys. Rev. D {\bf 49} (1994) 6587, gr-qc/9312023; 
``{\em
Increase of Black Hole Entropy in Higher Curvature Gravity,}'' Phys. Rev. D {\bf 52} (1995) 3518, gr-qc/9503020.

\bibitem{iyer}
V. Iyer and R.M. Wald, ``{\em Some Properties of Noether Charge and a Proposal for Dynamical Black Hole Entropy,}''
Phys. Rev. D {\bf 50} (1994) 846, gr-qc/9403028.

\bibitem{exotic}
M.-I. Park,
``{\em
Thermodynamics of exotic black holes, negative temperature, and Bekenstein-Hawking entropy,}''
Phys. Lett. B {\bf 647} (2007) 472, hep-th/0602114;
``{\em
Can Hawking temperatures be negative?,}''
Phys. Lett.  B {\bf 663 } (2008) 259, hep-th/0610140.


\bibitem{gurarie}
V. Gurarie, ``{\em Logarithmic operators in conformal field theory,}''
Nucl. Phys. B {\bf 410} (1993) 535, hep-th/9303160.

\bibitem{flohr}
M.A.I. Flohr, ``{\em On Modular Invariant Partition Functions of Conformal Field Theories with Logarithmic Operators,}''
Int.\ J.\ Mod.\ Phys.\ A {\bf 11} (1996) 4147, hep-th/9509166.















\end{thebibliography}
\end{document}